\documentclass[11pt]{article}
\usepackage{amsmath,amssymb,color,epsfig,cite}
\usepackage{graphicx}
\usepackage{setspace}

\usepackage{mathrsfs}

\usepackage{amsfonts}
\newcommand{\hoch}[1]{$\, ^{#1}$}

\makeatletter
\@addtoreset{equation}{section}
\makeatother

\newcommand{\be}{\begin{equation}}
\newcommand{\ee}{\end{equation}}
\newcommand{\bea}{\setlength\arraycolsep{2pt} \begin{eqnarray}}
\newcommand{\eea}{\end{eqnarray}}

\newcommand{\half}{{\textstyle{\frac{1}{2}}}}

\def\ndelta{\delta\hspace{-0.50em}\slash\hspace{-0.05em} }

\def\ft#1#2{{\textstyle{\frac{\scriptstyle #1}{\scriptstyle #2} } }}
\def\fft#1#2{{\frac{#1}{#2}}}

\def\0{{\sst{(0)}}}
\def\1{{\sst{(1)}}}
\def\2{{\sst{(2)}}}
\def\3{{\sst{(3)}}}
\def\4{{\sst{(4)}}}
\def\5{{\sst{(5)}}}
\def\6{{\sst{(6)}}}
\def\7{{\sst{(7)}}}
\def\8{{\sst{(8)}}}
\def\sst#1{{\scriptscriptstyle #1}}

\begin{document}

\begin{flushright}
\hfill{\hfill{MI-TH-1814}}

\end{flushright}

\vspace{15pt}
\begin{center}
{\Large {\bf Tower of subleading dual BMS charges}}

\vspace{15pt}
{\bf Hadi Godazgar\hoch{1}, Mahdi Godazgar\hoch{2} and 
C.N. Pope\hoch{3,4}}

\vspace{10pt}

\hoch{1} {\it Max-Planck-Institut f\"ur Gravitationsphysik (Albert-Einstein-Institut), \\
M\"uhlenberg 1, D-14476 Potsdam, Germany.}

\vspace{10pt}

\hoch{2} {\it Institut f\"ur Theoretische Physik,\\
Eidgen\"ossische Technische Hochschule Z\"urich, \\
Wolfgang-Pauli-Strasse 27, 8093 Z\"urich, Switzerland.}

\vspace{10pt}

\hoch{3} {\it George P. \& Cynthia Woods Mitchell  Institute
for Fundamental Physics and Astronomy,\\
Texas A\&M University, College Station, TX 77843, USA.}

\vspace{10pt}

\hoch{4}{\it DAMTP, Centre for Mathematical Sciences,\\
 Cambridge University, Wilberforce Road, Cambridge CB3 OWA, UK.}

 \vspace{15pt}
December 17, 2018

\vspace{20pt}

\underline{ABSTRACT}
\end{center}

\noindent
We supplement the recently found dual gravitational charges with dual charges for the whole BMS symmetry algebra. Furthermore, we extend the dual charges away from null infinity, defining subleading dual charges. These subleading dual charges complement the subleading BMS charges in the literature and together account for all the Newman-Penrose charges.    

\noindent

\thispagestyle{empty}

\vfill
E-mails: hadi.godazgar@aei.mpg.de, godazgar@phys.ethz.ch, pope@physics.tamu.edu

\pagebreak

\section{Introduction}

In recent work \cite{fakenews,dual0}, undertaken with the aim of providing a clear and explicit relation between the asymptotic BMS symmetry and gravitational charges of asymptotically flat spacetimes, we 
generalised the notion of BMS charges, as defined by Barnich and Troessaert 
\cite{BarTro}, in two complementary ways.  The Barnich-Troessaert BMS charges are derived from the general prescription of Barnich-Brandt \cite{BB} for defining asymptotic charges~\footnote{There exists an equivalent formalism for defining asymptotic charges, developed by Wald and collaborators \cite{IW, WZ}.  Here, we shall continue to work in the framework of the Barnich-Brandt formalism.}, which in this case turns out to be given by the integral of the Hodge dual of
a 2-form $H$
over a 2-sphere at null infinity:~\footnote{For an explicit expression for 
$H$, see equation \eqref{H}.} 
\begin{equation} \label{intro:BT}
 \ndelta \mathcal{Q}_{0}[\xi,g, \delta g] =\frac{1}{8 \pi G} \lim_{r\rightarrow\infty}  \int_{S}\, \star H[\xi,g, \delta g],
\end{equation}
where $\xi$ is the asymptotic symmetry generator, $g$ is the background metric and $\delta g$ is its variation.  The variation symbol 
$\ndelta$ denotes the fact that the charge is not necessarily integrable.  This is generally due to gravitational flux at null infinity.

In Ref.\ \cite{fakenews}, we extended the notion of BMS charges by defining 
subleading BMS charges as a $1/r$ expansion of the general prescription of 
Barnich-Brandt \cite{BB}, so that~\footnote{In principle, assuming 
analyticity, the tower of charges is infinite, with a charge at each
order in the $1/r$ expansion.  However, for 
physical reasons the metric expansion may only be analytic up to a 
certain order in $1/r$ (see e.g.\ Refs.\ \cite{Damour:1985cm,christ}). 
The tower of 
BMS charges will then naturally truncate at some corresponding order. \label{footnoteana}}
\begin{equation} \label{intro:1}
 \ndelta \mathcal{Q} =  \ndelta \mathcal{Q}_{0} +  \frac{\ndelta \mathcal{Q}_{1}}{r} +   \frac{\ndelta \mathcal{Q}_{2}}{r^2} +  
\frac{\ndelta \mathcal{Q}_{3}}{r^3} + \ldots\ ,
\end{equation}
and we showed that the $O(1/r^3)$ term gives 
five of the ten non-linear Newman-Penrose (NP) charges \cite{NP}.  Writing these charges in 
the Newman-Penrose formalism \cite{NP61}, we found that these five 
components, derived from the generalised BMS charge, correspond in some sense 
to the \emph{real} part of the NP charges.  An obvious question, then, 
is how do the other five \emph{imaginary} parts of the NP charges fit into 
this understanding of generalised BMS charges?  

In Ref.\ \cite{dual0}, inspired by the situation for 
electromagnetism \cite{Campiglia:2018dyi, Conde:2016csj}, which allows for 
\emph{electric} as well as \emph{magnetic} charges, we defined 
new \emph{dual} gravitational charges, associated with supertranslations, as the integral of the Barnich-Brandt
2-form $H$ itself (as opposed to its dual):
\begin{equation}\label{intro:dual}
 \ndelta \widetilde{\mathcal{Q}}_{0}[\xi,g, \delta g] =\frac{1}{8 \pi G} \lim_{r\rightarrow\infty}  \int_{S}\, H[\xi,g, \delta g].
\end{equation}
Together, $\ndelta \mathcal{Q}_{0}$ and $\ndelta \widetilde{\mathcal{Q}}_{0}$ 
can be viewed as the real and imaginary parts of a leading-order supertranslation charge, 
which can succinctly be written in terms of the leading-order terms in a 
$1/r$ expansion of the Newman-Penrose scalar $\Psi_2$, which is a
certain null-frame component of the Weyl tensor, and $\sigma$, which 
parametrises the shear 
of the null congruence of $\ell = \partial/\partial r$.  Thus, there is 
an attractive correspondence between, on the one hand,  
the real and imaginary parts of charges 
written in the complex Newman-Penrose formalism, and on the other hand 
``electric'' and ``magnetic'' (or ``dual'') BMS charges 
defined \textit{\`a la} Barnich-Brandt. 

  In this paper, we shall address the problem of generalising the 
dual charge 
$\ndelta \widetilde{\mathcal{Q}}_{0}$ of Ref.\ \cite{dual0} to a tower of
dual BMS charges (both $SL(2,\mathbb{C})$ and 
supertranslation charges) as a series in powers of $1/r$, in the same manner as 
the Barnich-Troessaert charge $\ndelta \mathcal{Q}_{0}$ was generalised 
in Ref.\ \cite{fakenews}.  The reason why in Ref.\ \cite{dual0} we were able 
to construct the dual charge at infinity (i.e.~at the order $1/r^0$) by
integrating the Barnich-Brandt 2-form $H$ as in equation (\ref{intro:dual})
was that if one takes $\delta g$ to be given by the action of the 
supertranslation 
generators, then $\ndelta \widetilde{\mathcal{Q}}_{0}[\xi,g, \delta_\xi g]$ 
defined in (\ref{intro:dual}) vanishes on-shell.  
However, beyond the
leading order, and including the SL$(2,\mathbb{C})$ part of the BMS group, it was established in Ref.\ \cite{dual0} 
that the corresponding variations of the subleading terms in 
the $1/r$ expansion of
the integral of $H$ do not vanish on-shell, and thus one does not 
get \textit{bona fide} subleading charges by this means.

 The question that we shall now address here is how does one generalise 
the dual charge \eqref{intro:dual} to the full BMS group and to subleading orders in a $1/r$ expansion 
away from null infinity?  Such a construction should provide 
an answer to the question raised by the results of Ref.\ \cite{fakenews}, 
i.e.\ it should presumably explain how the other five 
\emph{imaginary} parts of the NP charges come about.

We shall construct a tower of \textit{bona fide} dual 
gravitational BMS charges as a $1/r$ expansion away from null infinity, and 
we shall show that this does, in particular, give rise at the order $1/r^3$
to the five \emph{imaginary} parts of the NP charges.  
The tower of dual charges is given in terms of 
a new 2-form $\widetilde{H}$, such that
\begin{equation} \label{intro:dualG}
  \ndelta \widetilde{\mathcal{Q}}[\xi,g, \delta g] =\frac{1}{8 \pi G} 
 \int_{S}\, \widetilde{H}[\xi,g, \delta g]  =  
\ndelta \widetilde{\mathcal{Q}}_{0} +  
\frac{\ndelta \widetilde{\mathcal{Q}}_{1}}{r} +   
\frac{\ndelta \widetilde{\mathcal{Q}}_2}{r^2} +  
\frac{\ndelta \widetilde{\mathcal{Q}}_3}{r^3} + \ldots\ .
\end{equation}
Crucially, we construct $\widetilde H$ by requiring that the
integral in (\ref{intro:dualG}) should vanish on-shell when $\delta g$ is
taken to be given by the action of the BMS asymptotic symmetry generators, i.e.\ ${\ndelta \widetilde{\mathcal{Q}}[\xi, g, \mathcal{L}_{\xi} g]} = 0.$ It turns out that this condition uniquely defines $\widetilde{H}$.  Moreover, the more general condition that the central extension must be antisymmetric \cite{BB}, i.e.\  $\ndelta \widetilde{\mathcal{Q}}[\xi, g, \mathcal{L}_{\zeta} g] = - \ndelta \widetilde{\mathcal{Q}}[\zeta, g, \mathcal{L}_{\xi} g]$ is also satisfied.  Furthermore, properties of the BMS group ensure the existence of a charge algebra \cite{BB}.
The 2-form $\widetilde H$ that we find turns out 
to be equal to the Barnich-Brandt 2-form $H$ 
only at leading order and only for supertranslations
\begin{equation}
  \lim_{r\rightarrow\infty} (\widetilde{H} - H) = 0.
\end{equation}
Thus (\ref{intro:dualG}) gives the same leading-order result that we found
in Ref.\ \cite{dual0}, but now, we are able to extend the construction of
dual gravitational charges to the full BMS group and to all subleading orders in a $1/r$ expansion.

In section \ref{sec:pre}, we give some preliminary prerequisite information 
regarding asymptotically flat spacetimes.  For a more detailed exposition of
the notations and conventions we are using here, 
the reader is referred to section 2 of Ref.\ \cite{fakenews}.  
In section \ref{sec:dual}, we define the dual gravitational charge 
corresponding to the full BMS group. We find the full dual BMS charge at leading order in section \ref{sec:dual0Y} and investigate the 
dual charges associated with supertranslations up to order $1/r^3$ in a $1/r$ expansion in section \ref{sec:dualS}.  The results of 
this section are analogous to those obtained for the subleading BMS charges 
defined in Ref.\ \cite{fakenews}.  Perhaps, most significantly, in 
section \ref{sec:tI2} we find that the dual charge at order 
$1/r^3$ gives the imaginary parts of the NP charges, something that 
was missing in the analysis of Ref.\ \cite{fakenews}.  We finish with some 
comments in section \ref{sec:dis}.

\section{Preliminaries} \label{sec:pre}

We define asymptotically flat spacetimes 
as a class of spacetimes for which Bondi coordinates 
$(u,r,x^I=\{\theta, \phi\})$ may be defined, such that the metric takes 
the form \cite{bondi, sachs}
\begin{equation} \label{AF}
 d s^2 = - F e^{2 \beta} du^2 - 2 e^{2 \beta} du dr + 
r^2 h_{IJ} \, (dx^I - C^I du) (dx^J - C^J du),
\end{equation}
with the metric functions satisfying the following fall-off conditions~\footnote{We require even weaker fall-off conditions for leading dual BMS charges, \emph{viz}.\ $F = 1+ o(r^{0})$ and $\beta= o(r^{-1})$. However, we choose fall-off conditions such that we have both dual and Barnich-Troessaert BMS charges at leading order. In section~\ref{sec:dualS}, we impose stronger conditions in order to allow for NP charges.} at 
large $r$
\begin{align}
 F(u,r,x^I) &= 1 + \frac{F_0(u,x^I)}{r} +  o(r^{-1}), \notag \\[2mm]
 \beta(u,r,x^I) &= \frac{\beta_0(u,x^I)}{r^2} + o(r^{-2}), \notag \\[2mm] 
 C^I(u,r,x^I) &= \frac{C_0^I(u,x^I)}{r^2} + \frac{\log{r}}{r^3} D_{J} B^{IJ} +\frac{C_1^I(u,x^I)}{r^3} + o(r^{-3}), \notag \\[2mm] \label{met:falloff}
 h_{IJ}(u,r,x^I) &= \omega_{IJ} + \frac{C_{IJ}(u,x^I)}{r} + o(r^{-1}),
\end{align}
where $D_I$ is the standard covariant derivative associated with the unit round-sphere metric $\omega_{IJ}$ with coordinates $x^I=\{\theta, \phi\}$ on the 2-sphere. $B^{IJ}$ and $C_{IJ}$ are symmetric tensors with indices raised/lowered with the (inverse) metric on the 2-sphere.  Moreover, a residual gauge freedom 
allows us to require that
\begin{equation} \label{det:h}
 h =\omega,
\end{equation}
where $h \equiv \textup{det}(h_{IJ})$ and $\omega 
                      \equiv \textup{det}(\omega_{IJ}) =\sin\theta$.
                      
We assume, furthermore, that the components $T_{00}$ and $T_{0m}$ of the 
energy-momentum tensor in the null frame
fall off as
\begin{equation} \label{falloff:matter}
 T_{00} = o(r^{-4}), \qquad T_{0m} = o(r^{-3}).
\end{equation}
The Einstein equation then implies that
\begin{align}
  G_{00} = o(r^{-4}) &\quad \implies \quad \beta_0 = -\frac{1}{32}\, C^2,  \\
  G_{0m} = o(r^{-3}) &\quad \implies \quad C_0^I = -\half D_J C^{IJ},
\end{align}
where $C^2 \equiv C_{IJ} C^{IJ}$.

The asymptotic symmetry group corresponding to asymptotically flat spacetimes is the BMS group \cite{bondi, sachs}, whose corresponding algebra is generated by
\begin{gather}
 \xi = f \partial_u +  \xi^I  \partial_I - \frac{r}{2} \left( D_I \xi^I - C^I D_I f \right) \partial_r, \label{BMS} 
\end{gather}
where
\begin{equation}
  \xi^I= Y^I - \int_{r}^{\infty} dr' \frac{e^{2\beta}}{r'^2} h^{IJ} D_{J} f, \qquad f = s + \frac{u}{2} D_I Y^I.  \label{f}
\end{equation}
The $Y^I$ are the set of conformal Killing vectors on the round unit 2-sphere,
obeying 
\begin{equation} \label{Yeq}
 D_{(I} Y_{J)} = \frac{1}{2} D_K Y^K \omega_{IJ},
\end{equation}
and $s(x^I)$ is an arbitrary function that depends only on the angular coordinates and generates angle-dependent translations in the $u$-direction, 
which are called supertranslations ($\textup{ST}$).  Thus, 
\begin{equation}
\textup{BMS} = \textup{SL$(2,\mathbb{C})$} \ltimes \textup{ST}.
\end{equation}
The Abelian part of the algebra, generated by the supertranslations, is
\begin{equation} \label{BMSgen}
 \xi = s\, \partial_u -   \int_{r}^{\infty} dr' \frac{e^{2\beta}}{r'^2} h^{IJ} D_{J} s \  \partial_I - \frac{r}{2} \left( D_I \xi^I - C^I D _I s \right) \partial_r.
\end{equation}

As will become clear in what follows, it will be convenient to define twisted/dualised objects, as follows: for some symmetric tensor $X_{IJ}$, we define its \emph{trace-free} twist/dual by
\begin{equation} \label{Xtwist}
\widetilde{X}^{IJ} = X_K{}^{(I} \epsilon^{J)K}, \qquad \epsilon_{IJ} =  
\begin{pmatrix}                                                                                0 & 1 \\ -1 & 0                                                              \end{pmatrix} \sin \theta.
\end{equation}
If $X_{IJ}$ is, furthermore, trace-free, i.e.\ $\omega^{IJ} X_{IJ} = 0,$ 
then $X_K{}^{[I} \epsilon^{J]K} = 0$, so $\widetilde X^{IJ}$ is symmetric without the need for explicit symmetrisation
and we can simply write
\begin{equation}
 \widetilde{X}^{IJ} = X_K{}^{I} \epsilon^{JK}.
\end{equation}
If  $X$ and $Y$ are two symmetric trace-free tensors, then
\begin{equation} \label{teq}
 X_{IK} \widetilde{Y}^{JK} = - \widetilde{X}_{IK} Y^{JK}.
\end{equation}
Furthermore, if either one of the symmetric tensors $X$ or $Y$ is trace-free, then 
\begin{equation} \label{teqtr}
 X_{IJ} \widetilde{Y}^{IJ} = - \widetilde{X}_{IJ} Y^{IJ}.
\end{equation}

\section{Dual BMS charges} \label{sec:dual}

We define the dual BMS charge to be
\begin{equation} \label{Dcharge}
  \ndelta \widetilde{\mathcal{Q}}[\xi,g, \delta g] =\frac{1}{8 \pi G} \int_{S}\, \widetilde{H}[\xi,g, \delta g] = \frac{1}{8 \pi G} \int_{S} d \Omega \ \frac{\widetilde{H}_{\theta \phi}}{\sin \theta},
\end{equation}
where
\begin{equation} \label{tH}
 \widetilde{H} = \frac{1}{2} \Big\{\xi^c \nabla_b \delta g_{ac} - \frac{1}{2} \delta g_{bc} (\nabla_a \xi^c - \nabla^c \xi_a) \Big\} dx^a \wedge dx^b.
\end{equation}
This may be compared with the Barnich-Brandt 2-form $H$: \cite{BB}
\begin{equation} \label{H}
 H = \frac{1}{2} \Big\{ \xi_b g^{cd} \nabla_a \delta g_{cd} -\xi_b \nabla^c \delta g_{ac} +\xi^c \nabla_b \delta g_{ac} + \frac{1}{2} g^{cd} \delta g_{cd} \nabla_b \xi_a + \frac{1}{2} \delta g_{bc} (\nabla_a \xi^c - \nabla^c \xi_a) \Big\} dx^a \wedge dx^b.
\end{equation}
We found a unique expression for $\widetilde H$ by parameterising the most general possible covariant
2-form, built from terms bilinear in $\xi$ and $\delta g$ and involving one 
covariant derivative, and determining the constant coefficients by requiring
that its integral (\ref{Dcharge}) should vanish on-shell when $\delta g$ is
given by the action of an asymptotic symmetry generator,
i.e.
${\ndelta \widetilde{\mathcal{Q}}[\xi, g, \mathcal{L}_\xi g] = 0},$ 
where
\begin{equation}
\mathcal{L}_\xi g_{ab} = 2 \nabla_{(a} \xi_{b)}.\label{deltag}
\end{equation}
(Essentially, this amounted to putting arbitrary coefficients for the terms 
in the
expression (\ref{H}) of the Barnich-Brandt 2-form, and solving for them
by imposing the on-shell vanishing requirement.)

The general expression for the variation of a quantity $\mathcal{Q}$ is
\begin{equation}
\ndelta \mathcal{Q} = \delta \mathcal{Q}^{(int)} + \mathcal{N},
\end{equation}
where $\delta \mathcal{Q}^{(int)}$ is the integrable part, i.e.\ the ``time derivative", while the non-integrable term $\mathcal{N}$ quantifies the flux out of the system.  If ${\ndelta \mathcal{Q}[\xi, g, \mathcal{L}_\xi g]}=0$ on-shell, then we have a continuity equation and hence a charge corresponding to that asymptotic symmetry generator.

This reflects the fact that an asymptotic charge is not necessarily exactly conserved, viz.\ its time derivative is not necessarily zero, because of the existence of flux out of the system. Therefore, in this context we can define a charge if the quantity satisfies the analogue of a continuity equation, i.e. the charge changes by an amount given by the flux flowing out of the system. 
In appendix \ref{app:charge}, we show that ${\ndelta \widetilde{\mathcal{Q}}[\xi, g, \mathcal{L}_\xi g]}=0$, giving rise to a charge $\widetilde{\mathcal{Q}}^{(int)}$.  Furthermore, in appendix \ref{app:Htilde}, we verify that $\ndelta \widetilde{\mathcal{Q}}[\xi, g, \mathcal{L}_{\zeta} g] = - \ndelta \widetilde{\mathcal{Q}}[\zeta, g, \mathcal{L}_{\xi} g]$.  This together with the fact that the asymptotic symmetry generators belong to the BMS group implies that the charges defined by the asymptotic symmetry generators belong to a charge algebra \cite{BB}~\footnote{See, in particular, Theorem 4 of Ref.\ \cite{BB}.  Note that this theorem is labelled Theorem 2 in the published version.  For earlier results on this, in the context of the canonical formalism, we refer the reader to Ref.\ \cite{Brown:1986ed}.  We postpone a detailed study of the dual charge algebra to a future work.}. 

In fact all the features of the new dual charges we have introduced in this
paper are precisely analogous to those one encounters for the standard
BMS charges, defined from the Barnich-Brandt 2-form.  For example, at 
the leading $1/r^0$ order  
the variation $\ndelta \mathcal{Q}$ for the standard BMS charges also is non-integrable 
in general, owing to the presence of the Bondi news term $\mathcal{N}$ \cite{BarTro}. One
identifies the integrable part $\delta \mathcal{Q}^{(int)}$ of the variation
as defining the BMS charge $\mathcal{Q}^{(int)}$ at infinity; it is conserved
if the Bondi news vanishes.

Regarded as a $1/r$ expansion away from null infinity, we have~\footnote{See footnote \ref{footnoteana}.}
\begin{equation} \label{Dcharge:gen}
 \ndelta \widetilde{\mathcal{Q}}[\xi, g,\delta g]= \frac{1}{16 \pi G} \int_{S}\, d\Omega\ \Big\{ \ndelta \widetilde{\mathcal{I}}_0 + \frac{\ndelta \widetilde{\mathcal{I}}_1}{r} + \frac{\ndelta \widetilde{\mathcal{I}}_2}{r^2} + \frac{\ndelta \widetilde{\mathcal{I}}_3}{r^3} + o(r^{-3}) \Big\}.
\end{equation}
Hence, we find a tower of dual charges, which can be viewed as the charges dual to the BMS charges found in Ref.\ \cite{fakenews}
\begin{equation} \label{BMS:gen}
 \ndelta \mathcal{Q}[\xi, g,\delta g]=\frac{1}{8 \pi G}  \int_{S}\, \star H[\xi,g, \delta g]= \frac{1}{16 \pi G} \int_{S}\, d\Omega\ \Big\{ \ndelta \mathcal{I}_0 + \frac{\ndelta \mathcal{I}_1}{r} + \frac{\ndelta \mathcal{I}_2}{r^2} + \frac{\ndelta \mathcal{I}_3}{r^3} + o(r^{-3}) \Big\}.
\end{equation}
In particular, as we shall demonstrate in section \ref{sec:tI3}, 
$\widetilde{\mathcal{I}}_3$ gives the five complementary NP charges \cite{NP}
that were missing in the analysis of Ref.\ \cite{fakenews}.

We proceed to describe the leading-order dual charge for the full BMS group, before we investigate the subleading terms, corresponding to supertranslations only, in $\ndelta \widetilde{\mathcal{Q}}$ in the $1/r$ expansion given in equation \eqref{Dcharge:gen}.

\section{Leading-order dual charges} \label{sec:dual0Y}
In this section we derive the leading-order dual BMS charges.  These charges ought to be viewed as duals of the Barnich-Troessaert charges found in Ref.\ \cite{BarTro}.  The dual charge as defined in Ref.\ \cite{dual0} by taking the integral of the Barnich-Brandt 2-form $H$ (as opposed to the integral of its Hodge dual) cannot incorporate the SL$(2,\mathbb{C})$ part of the BMS group.  However, the dual charge defined in section \ref{sec:dual} gives a charge for the full BMS group, as argued for in appendix \ref{app:charge}.  

Using the definition of the asymptotic symmetry generator, given by equation \eqref{BMS} and the metric coefficients defined in equations \eqref{met:falloff}, it is relatively simple to show that
\begin{align}
 \frac{\widetilde{H}_{\theta \phi}}{\sin \theta} &= \frac{1}{2} \epsilon^{IJ} H_{IJ} \notag \\
                                             &= \frac{1}{2} \epsilon^{IJ} \Big[\xi^c \nabla_J \delta g_{Ic} - \frac{1}{2} \delta g_{Jc} (\nabla_I \xi^c - \nabla^c \xi_I) \Big] \notag \\
                                             &= \frac{1}{2} \epsilon^{IJ} \Bigg[ r \Big( Y^K D_J \delta C_{IK} - \frac{1}{2} \delta C_{JK} (D_I Y^K - D^K Y_I) \Big) + D_I(f \delta C_{0 J}) - \frac{1}{4} D_{I}(Y_J \delta C^2)\notag \\
                                             & \hspace{12mm} + D^K f\, D_I \delta C_{JK} + \frac{1}{2} f \partial_u C_{IK} \delta C_J{}^K + \frac{1}{2} Y^L \delta C_{JK} D_L C_{I}{}^K + \frac{1}{4} D_L Y^L C_{IK} \delta C_{J}{}^K \notag \\
                                             & \hspace{12mm}  + \frac{1}{2} C_{IL} \delta C_{JK} D^K Y^L - \frac{1}{2} C^{KL} \delta C_{JK} D_L Y_I  \Bigg] + O(1/r).
\end{align}
Ignoring total derivatives, as these will integrate to zero, and freely integrating by parts, the expression above simplifies to
\begin{align}
 \frac{\widetilde{H}_{\theta \phi}}{\sin \theta} &= \frac{1}{2} \epsilon^{IJ} \Bigg[ \frac{1}{2} r \delta C_{JK} (D_I Y^K + D^K Y_I) - f D^K D_I \delta C_{JK} + \frac{1}{2} f \partial_u C_{IK} \delta C_J{}^K  \notag \\
                                             & \hspace{12mm}  + \frac{1}{4} Y^L \delta C_{J}{}^{K} D_L C_{IK} - \frac{1}{4} Y^L C_{IK}  D_L\delta C_{J}{}^K \notag \\
                                             & \hspace{12mm}  + \frac{1}{2} C_{IL} \delta C_{JK} D^K Y^L - \frac{1}{2} C^{KL} \delta C_{JK} D_L Y_I  \Bigg] + O(1/r) \notag \\
                                             &= \frac{1}{2} \Bigg[  r \delta \widetilde{C}^{IJ} D_{(I} Y_{J)} - f D_I D_J \delta \widetilde{C}^{IJ} + \frac{1}{2} f \partial_u C_{IJ} \delta \widetilde{C}^{IJ}  + \frac{1}{4} Y^K \delta (\widetilde{C}^{IJ} D_K C_{IJ}) \notag \\
                                             & \hspace{12mm} + \frac{1}{2} \big( C_{IK} \delta \widetilde{C}^{JK} - C^{JK} \delta \widetilde{C}_{IK} \big) D_J Y^I  \Bigg] + O(1/r).  \label{tH:12}
\end{align}
Now, using equation \eqref{Yeq} and the fact that $\delta \widetilde{C}^{IJ}$ is trace-free implies that the order $r$ terms in the expression above vanish, as they should.  Moreover, given that 
\begin{equation}
 C_{IK} \delta \widetilde{C}_J{}^{K} = \frac{1}{2} C_{KL} \delta \widetilde{C}^{KL} \omega_{IJ} - \frac{1}{4} \delta C^2 \epsilon_{IJ},
\end{equation}
which can simply be derived from observing that the symmetric and antisymmetric parts of the expression on the left hand side of the above equation must be proportional to $\omega_{IJ}$ and $\epsilon_{IJ}$, respectively,
the expression for $\widetilde{H}_{\theta \phi}$ in equation \eqref{tH:12} simplifies to
\begin{equation}
 \frac{\widetilde{H}_{\theta \phi}}{\sin \theta} = \frac{1}{2} \Bigg[ - f D_I D_J \delta \widetilde{C}^{IJ} + \frac{1}{2} f \partial_u C_{IJ} \delta \widetilde{C}^{IJ}  + \frac{1}{4} Y^K \delta (\widetilde{C}^{IJ} D_K C_{IJ}) 
 + \frac{1}{2} D_I \widetilde{Y}^I \delta C^2 \Bigg] + O(1/r), 
\end{equation}
where
\begin{equation}
 \widetilde{Y}^I = \epsilon^{IJ} Y_J.
\end{equation}
In summary, we find that
\begin{align}
 \ndelta \widetilde{\mathcal{Q}}_0 =\frac{1}{16 \pi G} \int_{S}\, \Bigg[ \delta \Bigg( - f D_I D_J \widetilde{C}^{IJ}  + \frac{1}{4} Y^K \widetilde{C}^{IJ} D_K C_{IJ} - \frac{1}{4} & \widetilde{Y}^I D_I  C^2 \Bigg) \notag\\
                                                        &\qquad + \frac{1}{2} f \partial_u C_{IJ} \delta \widetilde{C}^{IJ} \Bigg]. \label{dual0:Full}
\end{align}
This dual charge may be compared with the charge in \cite{BarTro}~\footnote{See equations (3.2) and (3.3) of Ref.\ \cite{BarTro} with the following translations in notation: $M=-1/2\, F_0$ and $N^I = -3/2 \, C_{1}^I$.}
\begin{align}
 \ndelta \mathcal{Q}_0 =\frac{1}{16 \pi G} \int_{S}\, \Bigg[ \delta \Bigg( - 2 f F_0  + Y^K \Big[ -3 C_{1 K} + \frac{1}{16}   D_K  C^2 + C_{JK} D_I C^{IJ} \Big]& \Bigg) \notag\\
                                                        &\quad + \frac{1}{2} f \partial_u C_{IJ} \delta C^{IJ} \Bigg].
\end{align}

\section{Subleading dual charges} \label{sec:dualS}
In the previous section, we computed the leading-order dual BMS charge for the full BMS group.  In this section, for simplicity, we restrict ourselves to the most distinctive part of the BMS group, given by supertranslations, and compute the subleading charges.  Thus, hereafter, the generators that will be of interest are those given by equation \eqref{BMSgen}.  

Furthermore, in this section we require stronger fall-off conditions:
\begin{align}
 F(u,r,x^I) &= 1 + \frac{F_0(u,x^I)}{r} + \frac{F_1(u,x^I)}{r^2} + \frac{F_2(u,x^I)}{r^3} + \frac{F_3(u,x^I)}{r^4} + o(r^{-4}), \notag \\[2mm]
 \beta(u,r,x^I) &= \frac{\beta_0(u,x^I)}{r^2} + \frac{\beta_1(u,x^I)}{r^3} + \frac{\beta_2(u,x^I)}{r^4} + o(r^{-4}), \notag \\[2mm] 
 C^I(u,r,x^I) &= \frac{C_0^I(u,x^I)}{r^2} + \frac{C_1^I(u,x^I)}{r^3} + \frac{C_2^I(u,x^I)}{r^4} + \frac{C_3^I(u,x^I)}{r^5} + o(r^{-5}), \notag \\[2mm] \label{met:falloff2}
 h_{IJ}(u,r,x^I) &= \omega_{IJ} + \frac{C_{IJ}(u,x^I)}{r} + \frac{C^2 \omega_{IJ}}{4 r^2} + \frac{D_{IJ}(u,x^I)}{r^3} + \frac{E_{IJ}(u,x^I)}{r^4} + o(r^{-4}).
\end{align}
Further to the fall-off conditions \eqref{falloff:matter}, we require
\begin{equation} \label{falloff:matter2}
 T_{00} = o(r^{-5}), \qquad T_{0m} = o(r^{-3}),
\end{equation}
which implies
\begin{align}
  G_{00} = o(r^{-5}) &\quad \implies \quad \beta_0 = -\frac{1}{32}\, C^2, \quad \beta_1 = 0,  \\
  G_{0m} = o(r^{-3}) &\quad \implies \quad C_0^I = -\half D_J C^{IJ}.
\end{align}
These stronger fall-off conditions are needed for the existence of NP charges, whose origin we explain in terms of subleading BMS and dual BMS charges in this section. These conditions allow us to define charges up to order $1/r^3$. In order to define yet further higher order charges we need to impose even stronger fall-off conditions. 

\subsection{Dual charge at $O(r^{0})$} \label{sec:tI0}

For the leading-order charge, the contribution of the supertranslations can be simply deduced from the general result \eqref{dual0:Full} by turning off the SL$(2,\mathbb{C})$ generators, i.e.\ the $Y^I$'s.  Hence, from equation \eqref{f}, $f=s$ and charge \eqref{dual0:Full} reduces to
\begin{equation} \label{tI0}
 \ndelta \widetilde{\mathcal{I}}_0 = \delta \big( -  s D_I D_J \widetilde{C}^{IJ} 
\big) + \frac{s}{2} \partial_u C_{IJ} \delta \widetilde{C}^{IJ}.
\end{equation}
This agrees with the result of Ref.\ \cite{dual0} since at the leading order, the 2-forms $H$ and $\widetilde{H}$ coincide
\begin{equation} \label{HtH}
  \lim_{r\rightarrow\infty} (\widetilde{H} - H) = 0.
\end{equation}

As emphasised in Ref.\ \cite{dual0}, the leading-order dual charge 
is integrable if and only if
\begin{equation}
 \partial_u C_{IJ} = 0,
\end{equation}
i.e.\ the Bondi news vanishes.  Recall from section 3.1 of 
Ref.\ \cite{fakenews} that an equivalent statement holds for the 
Barnich-Troessaert charge $\ndelta \mathcal{I}_0$ \cite{BarTro}:
\begin{equation} \label{I0}
 \ndelta \mathcal{I}_0 = \delta \big( -2 s F_{0} \big) + \frac{s}{2} \partial_u C_{IJ} \delta C^{IJ}.
\end{equation}
Moreover, as discussed in Ref.\ \cite{dual0}, the integrable parts of the two sets of charges may be written as the real and imaginary parts of a single expression~\footnote{See section 4 of Ref.\ \cite{fakenews} for a brief introduction to the Newman-Penrose scalars.}
\begin{equation} \label{NPF:0}
 \mathcal{Q}_0 = - \frac{1}{4\pi G} \int d\Omega\ s\; 
	(\psi_2^0 + \sigma^0 \partial_u \bar{\sigma}^0).
\end{equation}
More precisely (recalling that $s$ is a real quantity),
\begin{equation}
 \mathcal{Q}_0 = \mathcal{Q}_0^{(int)} - i \widetilde{\mathcal{Q}}_0^{(int)},
\end{equation}
where, from equations \eqref{I0} and \eqref{tI0}, respectively,
\begin{equation}
 \mathcal{Q}_0^{(int)} = \frac{1}{16\pi G} \int d\Omega\  (-2 s F_{0}), 
\qquad \widetilde{\mathcal{Q}}_0^{(int)} =  
\frac{1}{16\pi G} \int d\Omega\ (- s D_I D_J \widetilde{C}^{IJ}).
\end{equation}

\subsection{Dual charge at $O(r^{-1})$} \label{sec:tI1}

At the next order a simple, if rather tedious, calculation shows that up to 
total derivatives, which will vanish under integration,
\begin{equation} \label{tI1}
 \ndelta \widetilde{\mathcal{I}}_1 = 0.
\end{equation}
Recall (see section 3.2 of Ref.\ \cite{fakenews}) that assuming 
\begin{equation} \label{T01}
 T_{01} = o(r^{-4}),
\end{equation}
the Einstein equation implies that
\begin{equation} \label{I1}
 \ndelta \mathcal{I}_1 = 0.
\end{equation}
More generally, without assuming equation \eqref{T01}, we may equivalently define
\begin{equation} \label{NPF:1}
 \mathcal{Q}_1 = - \frac{1}{8\pi G} \int d\Omega\ s\; (\psi_2^1 - \bar{\eth} \psi_1^0),
\end{equation}
since
\begin{equation}
 \Re(\psi_2^1 - \bar{\eth} \psi_1^0) = -\frac{1}{2} \mathcal{I}_1, \qquad \Im(\psi_2^1 - \bar{\eth} \psi_1^0) = 0.
\end{equation}

\subsection{Dual charge at $O(r^{-2})$} \label{sec:tI2}

A similar long, but simple, calculation finds that
\begin{align}
 \ndelta \widetilde{\mathcal{I}}_2 &=  s\  D_I D_J \delta \Big( -  \widetilde{D}^{IJ} + \frac{1}{16} \,  C^2 \widetilde{C}^{IJ}\Big) \notag \\[2mm]
  &  + s \Bigg( \frac{1}{2} \Big[ \partial_u D_{IJ} \delta \widetilde{C}^{IJ} - \delta D_{IJ} \partial_u \widetilde{C}^{IJ} \Big] - \frac{1}{16} C_{IJ} \Big[ \partial_u C^2  \delta \widetilde{C}^{IJ} - \delta C^2 \partial_u \widetilde{C}^{IJ} \Big] + D_I (C_{1 J} \delta \widetilde{C}^{IJ} ) \notag \\
  & \hspace{10mm} - \frac{1}{16} D_I (D_J C^2 \delta \widetilde{C}^{IJ}) - \frac{1}{2} D_{I} (C_{JK} D_L C^{KL} \delta \widetilde{C}^{IJ}) \Bigg). \label{Dcharge:tI2}
\end{align}
Note that this is very similar to $\ndelta\mathcal{I}_2$ 
(see equation (3.12) of Ref.\ \cite{fakenews}).  In particular, 
the integrable part of one is obtained by taking the twist of the tensor fields
in the other.  The non-integrable part provides an obstruction to 
the conservation of the 
integrable charge, and in analogy with the nomenclature adopted for the 
non-integrable parts of the charges at subleading order in 
Ref.\ \cite{fakenews}, we may describe such terms here 
as ``twisted fake news.''  

  In what follows, we consider whether $s$ can be appropriately chosen such 
that the twisted fake news vanishes, leaving a conserved/integrable charge.  
In order to proceed, we need to know how $C_{IJ}$ and $D_{IJ}$ transform 
under the action of the asymptotic symmetry group; these are given 
by equations (2.30) and (2.32) of Ref.\ \cite{fakenews}, which we reproduce here for 
convenience
 \begin{align}
 \delta C_{IJ} &= s \partial_u C_{IJ} + \Box s\ \omega_{IJ} - 2 D_{(I} D_{J)} s, \label{var:C} \\[2mm]
 \delta D_{IJ} &= s \partial_u D_{IJ} + \Big[ \frac{1}{16} C^2 \Box s - \frac{1}{16} D^K C^2 D_K s - \frac{1}{2} C^{LM} D^K C_{KL} D_Ms + C_1^K D_K s \Big] \omega_{IJ}  \notag \\
 & - 2 C_{1 \, (I} D_{J)}s- \frac{1}{4} C_{IJ} C^{KL} D_K D_L s  - \frac{1}{8}C^2 D_{I} D_{J}s + \frac{1}{8} D_{(I} C^2 D_{J)}s + D_K C^{KL} C_{L(I} D_{J)}s. \label{var:D}   
\end{align}
Moreover, assuming that $T_{mm} = o(r^{-4})$, we have \cite{fakenews}
\begin{align}
 \partial_u D_{IJ} =& \frac{1}{8} C_{IJ} \partial_u C^2 - \frac{1}{4} F_{0} C_{IJ}  - \frac{1}{2} D_{(I} C_{1\, J)} - \frac{1}{8} C_{IJ} D_K D_L C^{KL} \notag \\
            & + \frac{1}{32} D_I D_J C^2 + \frac{1}{2} D_{(I}(C_{J)K} D_L C^{KL}) - \frac{1}{8} D_I C^{KL} D_J C_{KL}    \notag \\
            & + \frac{1}{4} \omega_{IJ} \Big[ D_K C_1^K 
-\frac{5}{16} \Box C^2  + D^M C^{KL} \big( D_{K} C_{LM}- 
\frac{1}{4} D_{M} C_{KL}\big) + C^2 \Big]. \label{uD}
\end{align}
In order to simplify the analysis, we begin by noting that since there is no 
Einstein equation for $F_0$, terms involving $F_0$ in the non-integrable 
part of $\ndelta \widetilde{\mathcal{I}}_2$, given in the second and 
third lines of equation \eqref{Dcharge:tI2}, 
would have to vanish independently. Using the two equations 
\eqref{var:D} and \eqref{uD}, it is easy to see that
\begin{align}
 \ndelta \widetilde{\mathcal{I}}_2^{(non-int)}|_{F_0 \; \textrm{terms}} &= - \frac{1}{8} s F_0 C_{IJ} \Big[\delta \widetilde{C}^{IJ} - s \partial_u \widetilde{C}^{IJ}\Big] \notag \\
                                                                    &= \frac{1}{8} s F_0 \widetilde{C}_{IJ} \Big[\delta C^{IJ} - s \partial_u C^{IJ}\Big]  \notag \\
                                                                    &= - \frac{1}{4} s F_0 \widetilde{C}^{IJ} D_{I} D_{J} s,
\end{align}
where, in the second equality, we have used equation \eqref{teqtr} and, 
in the third equality, we have used \eqref{var:C}.  For arbitrary $F_0$ 
and $\widetilde{C}^{IJ}$ (which is trace-free), the above expression 
vanishes if and only if
\begin{equation} \label{tI2:s}
  D_I D_J s = \frac{1}{2} \omega_{IJ} \Box s.
\end{equation}
This is precisely the condition that $s$ is an 
$\ell=0$ or $\ell=1$ spherical harmonic (see appendix C of Ref.~\cite{fakenews}), 
which implies in particular that
\begin{equation}
 \delta C_{IJ} = s \partial_u C_{IJ}, 
\qquad \delta \widetilde{C}_{IJ} = s \partial_u \widetilde{C}_{IJ}.
\end{equation}
Proceeding, while taking $s$ to satisfy equation \eqref{tI2:s}, 
the non-integrable part of $\ndelta \widetilde{\mathcal{I}}_2$ simplifies to
\begin{align}
 \ndelta \widetilde{\mathcal{I}}^{(non-int)}_2 =& \Big( C_{1 I} D_{J}s - \frac{1}{16} D_{I} C^2 D_{J}s - \frac{1}{2} D_K C^{KL} C_{LI} D_{J}s \Big) \delta \widetilde{C}^{IJ}  \notag \\
                                         &+s D_I (C_{1 J} \delta \widetilde{C}^{IJ}) -  \frac{1}{16} s D_I (D_J C^2 \delta \widetilde{C}^{IJ}) - \frac{1}{2} s D_{I} (C_{JK} D_L C^{KL} \delta \widetilde{C}^{IJ}),
\end{align}
which clearly forms a total derivative, and thus vanishes under integration.  
We therefore conclude that if $s$ is an $\ell=0$ or $\ell=1$ spherical 
harmonic,
\begin{equation}
  \ndelta \widetilde{\mathcal{I}}_2^{(non-int)} = 0
\end{equation}
and so
\begin{equation}
 \widetilde{\mathcal{I}}_2 = s\   D_I D_J \Big( -  \widetilde{D}^{IJ} + \frac{1}{16} \,  C^2 \widetilde{C}^{IJ}\Big).
\end{equation}
However, up to total derivatives
\begin{equation}
 \widetilde{\mathcal{I}}_2 = D_I D_J s\   
\Big( -  \widetilde{D}^{IJ} + \frac{1}{16} \,  C^2 \widetilde{C}^{IJ}\Big),
\end{equation}
and this in fact vanishes upon use of equation \eqref{tI2:s}.  
This analysis is, rather remarkably, completely analogous to that of 
$\ndelta \mathcal{I}_2$ in Ref.~\cite{fakenews}: the non-integrable part can be 
made to vanish if and only if $s$ is an $\ell=0$ or $\ell=1$ spherical 
harmonic, in which case the integrable charge itself turns out to be trivial.

Finally, we express the integrable parts of 
$\ndelta \mathcal{I}_2$ and $\ndelta \widetilde{\mathcal{I}}_2$ as the real and 
imaginary parts, respectively, of a single charge written in terms of Newman-Penrose scalars.  Defining
\begin{equation} \label{NPF:2}
 \mathcal{Q}_2 =  \frac{1}{24\pi G} \int d\Omega\ s\; \bar{\eth}^2 \psi_0^0,
\end{equation}
we find that this complex quantity may be written in terms of the 
integrable charges as 
\begin{equation}
 \mathcal{Q}_2 = \mathcal{Q}_2^{(int)} - i \widetilde{\mathcal{Q}}_2^{(int)},
\end{equation}
where
\begin{gather}
 \mathcal{Q}_2^{(int)} = \frac{1}{16\pi G} \int d\Omega\  s\  D_I D_J  \Big( -  D^{IJ} + \frac{1}{16} \,  C^2 C^{IJ}\Big), \notag \\ 
 \widetilde{\mathcal{Q}}_2^{(int)} =  \frac{1}{16\pi G} \int d\Omega\ s\  D_I D_J \Big( -  \widetilde{D}^{IJ} + \frac{1}{16} \,  C^2 \widetilde{C}^{IJ}\Big).
\end{gather}

\subsection{Dual charge at $O(r^{-3})$} \label{sec:tI3}

Lastly, we consider the dual charge at order $1/r^3$.  We find after
some algebra that
\begin{align}
 \ndelta \widetilde{\mathcal{I}}_3 = - & s \ D_I D_J \delta \widetilde{E}^{IJ}  \notag\\[2mm]
  & \hspace{-10mm} + s \Bigg( \frac{1}{2} \Big[ \partial_u E_{IJ} \delta \widetilde{C}^{IJ} - \delta E_{IJ} \partial_u \widetilde{C}^{IJ} \Big] - \frac{1}{4} D^K (C_{1I} C_{JK} \delta \widetilde{C}^{IJ} ) \notag \\
  &+ D^I \left(\Big[ C_2^J - \frac{3}{4} \left( D_K D^{JK} - C^{JK} C_{1\, K} \right) -\frac{1}{64} C^2 D_K C^{JK} +\frac{1}{16} C^{JK} D_{K}C^2 \Big] \delta \widetilde{C}_{IJ} \right)   \notag \\
  &  + \frac{1}{4} D^K (\delta \widetilde{C}^{IJ}  D_I D_{JK}) - \frac{5}{4} D_I ( D_{JK}  D^K \delta \widetilde{C}^{IJ}) + \frac{1}{16} D^K ( \delta \widetilde{C}^{IJ} C_{JK} D_I C^2) \notag \\ 
  &   -\frac{5}{64}\Big[ \delta \widetilde{C}^{IJ} D^K (C^2 D_I C_{JK}) - C_{JK} D_I (C^2 D^K \delta \widetilde{C}^{IJ}) \Big] \Bigg). \label{Dcharge:tI3p}
\end{align}
Assuming that $T_{0m}=o(r^{-5})$, the Einstein equation gives an equation 
for $C_2^I$, equation (2.18) of Ref.\ \cite{fakenews}:
\begin{equation}
 C_2^I = \frac{3}{4} \left( D_J D^{IJ} - C^{IJ} C_{1\, J} \right) +\frac{1}{64} C^2 D_J C^{IJ} -\frac{1}{16} C^{IJ} D_{J}C^2.
\end{equation}
Substituting this equation into \eqref{Dcharge:tI3p} gives
\begin{align}
 \ndelta \widetilde{\mathcal{I}}_3 = - & s \ D_I D_J \delta \widetilde{E}^{IJ}  \notag\\[2mm]
  & \hspace{-10mm} + s \Bigg( \frac{1}{2} \Big[ \partial_u E_{IJ} \delta \widetilde{C}^{IJ} - \delta E_{IJ} \partial_u \widetilde{C}^{IJ} \Big] - \frac{1}{4} D^K (C_{1I} C_{JK} \delta \widetilde{C}^{IJ} ) \notag \\
  &  + \frac{1}{4} D^K (\delta \widetilde{C}^{IJ}  D_I D_{JK}) - \frac{5}{4} D_I ( D_{JK}  D^K \delta \widetilde{C}^{IJ}) + \frac{1}{16} D^K ( \delta \widetilde{C}^{IJ} C_{JK} D_I C^2) \notag \\ 
  &   -\frac{5}{64}\Big[ \delta \widetilde{C}^{IJ} D^K (C^2 D_I C_{JK}) - C_{JK} D_I (C^2 D^K \delta \widetilde{C}^{IJ}) \Big] \Bigg). \label{Dcharge:tI3}
\end{align}
Comparing this dual variation with the analogous term $\ndelta \mathcal{I}_3$ 
given by equation (3.28) of Ref.\ \cite{fakenews}, we find that they 
are very similar.  For the integrable parts, noting that we found (see equation (3.42) of Ref.\ \cite{fakenews})
\begin{equation} \label{I3:int}
 \mathcal{I}_3^{(int)} = 
s\, D_I D_J \Bigg( -E^{IJ} + \frac{1}{2}\, \textup{tr}E\ \omega^{IJ} \Bigg),
\end{equation}
we see that the integrable parts are related by replacing the tensor 
fields in one by the twists of the fields in the other. (Note that
$\textup{tr}\widetilde E=0$.)
  
As with $\ndelta \mathcal{I}_3$, there exist non-integrable terms also, 
and one may consider, as we did previously at order $1/r^2$, 
whether there exists 
some choice of the function $s$ such that the non-integrable part of 
$\ndelta \widetilde{\mathcal{I}}_3$, given by the last three lines of equation 
\eqref{Dcharge:tI3}, vanishes.  We turn to this consideration in what follows.
The variation of $E_{IJ}$ under the action of a supertranslation is 
given by \cite{fakenews}
 \begin{align}
  \delta E_{IJ} & = s \partial_u E_{IJ}  + \Big[ \frac{1}{4} D^{KL} D_K D_L s + \frac{3}{2} D_K D^{KL} D_L s  - \frac{5}{4}  C^{KL} C_{1\, K} D_L s - \frac{1}{64}  C^2 C^{KL} D_K D_L s  \notag \\ 
 & + \frac{3}{64} \Big( C^{KL} D_K C^2+ 2 C^2 D_K C^{KL} \Big) D_L s\Big] \omega_{IJ} + \frac{1}{2} C_{1\, (I} C_{J)K} D^K s - \frac{5}{2} D^K( D_{K(I} D_{J)} s) \notag \\ 
 & - \frac{1}{2} D^K s D_{(I} D_{J)K}  + \frac{5}{32} D^K (C^2 C_{K(I} D_{J)} s) + \frac{5}{32} C^2 D^K s D_{(I} C_{J)K} - \frac{1}{8} C_{K(I} D_{J)}C^2 D^K s. \label{var:E}
\end{align}
Also, assuming that $T_{mm} = o(r^{-5})$, the Einstein equation implies that \cite{fakenews}
\begin{align}
   \partial_u E_{IJ} =& \frac{1}{2} D^K(C_{1\, (I} C_{J)K}) - \frac{1}{2} D^K D_{(I} D_{J)K} + \frac{5}{32} D^K(C^2 D_{(I} C_{J)K})  \notag \\
            & - \frac{1}{8} D^K (C_{K(I} D_{J)} C^2) + \frac{1}{2} \omega_{IJ} \Big[ D^{KL} \partial_u C_{KL} - \frac{1}{4} C^2 F_0 - \frac{1}{2} C_{1}^{K} D^L C_{KL}    \notag \\            
            & - C^{KL} D_K C_{1\, L}+ \frac{1}{2} D^K D^L D_{KL} - \frac{1}{32} C^2 D^K D^L C_{KL} + \frac{5}{32} C^{KL} D_K D_L C^2  \notag \\
            & - \frac{1}{16} C_{KL} D_M C^{MK} D_N C^{NL} + \frac{3}{32} C^{KL} D_K C^{MN} D_L C_{MN} \Big].  \label{uE}
\end{align}
Rewriting
\begin{align}
 s  \Big[ \partial_u E_{IJ} \delta \widetilde{C}^{IJ} - \delta E_{IJ} \partial_u \widetilde{C}^{IJ} \Big] = - \Big( \delta E_{IJ} - s \partial_u E_{IJ} \Big)\delta  \widetilde{C}^{IJ} + \delta E_{IJ} \Big( \delta \widetilde{C}^{IJ} - s \partial_u \widetilde{C}^{IJ} \Big),
\end{align}
$\ndelta \widetilde{\mathcal{I}}^{(non-int)}_3$ simplifies to
\begin{align}
 \ndelta \widetilde{\mathcal{I}}^{(non-int)}_3 =& \frac{1}{4} D^K \Bigg[ \Big( - C_{1 I} C_{JK} + D_{I} D_{JK} + \frac{1}{4} C_{JK} D_I C^2 \Big) s\, \delta \widetilde{C}^{IJ} \Bigg] \notag \\
                                & + \frac{5}{4} \Bigg[ D^K (D_{JK} D_I s) \delta \widetilde{C}^{IJ} - s D_I (D_{JK} D^K \delta \widetilde{C}^{IJ}) \Bigg] \notag \\
                                & + \frac{5}{64} \Bigg[s C_{JK} D_I (C^2 D^K \delta \widetilde{C}^{IJ}) - D^K (C^2 D_I[s C_{JK}]) \delta \widetilde{C}^{IJ} \Bigg] \notag \\
                                & - \frac{1}{4}  D^K\widetilde{X}_{IJK} \Big( \delta C^{IJ} - s \partial_u C^{IJ} \Big), \label{tI3:non}
\end{align}
where
\begin{equation} \label{tX}
 \widetilde{X}_{IJK} = s\, C_{1 I} \widetilde{C}_{JK} - s\, D_{I} \widetilde{D}_{JK} - 5 \widetilde{D}_{JK} D_{I} s - \frac{1}{4} s\, \widetilde{C}_{JK} D_I C^2 + \frac{5}{16} C^2 D_I (s \widetilde{C}_{JK})
\end{equation}
and we have used equation \eqref{teqtr}.
Note that the expression in the first line of equation \eqref{tI3:non} is a 
total derivative, which will integrate to zero.  Moreover, integrating by 
parts and dropping total derivatives, the expressions on the second and 
third lines cancel.  This leaves the expression on the fourth line, which, 
using equation \eqref{var:C}, and integrating by parts, reduces to
\begin{equation}
 \ndelta \widetilde{\mathcal{I}}^{(non-int)}_3 = - \frac{1}{2} \widetilde{X}_{IJK} D^K \Big( D^I D^J s - \frac{1}{2} \omega^{IJ} \Box s  \Big).
\end{equation}
Note that $\widetilde{X}_{IJK}$ as defined in equation \eqref{tX} is symmetric 
and trace-free in its indices $(JK)$.  Thus, for arbitrary metric functions 
$C_1^I$, $C_{IJ}$ and $D_{IJ}$, the expression above vanishes if and only if
\begin{equation} \label{tI3:s}
 D^K \Big( D^I D^J s - \frac{1}{2} \omega^{IJ} \Box s  \Big) = \omega^{JK}\, 
  U^I + \epsilon^{JK}\, W^I,
\end{equation}
where the vectors $U^I$ and $W^I$ can be 
found by multiplying the above equation by $\omega_{JK}$ and $\epsilon_{JK}$, respectively.  
Only the symmetric part in $(JK)$ is relevant here since 
$\widetilde X_{IJK}$ is symmetric in $(JK)$. Using the fact 
that
\begin{equation}
 [D_I, D_J] V_K = R_{IJK}{}^L V_{L}, \qquad R_{IJKL} = \omega_{IK}\, \omega_{JL} - \omega_{IL}\, \omega_{JK},
\end{equation}
we find that the trace-free, symmetric $(JK)$ projection of the left hand side of
equation \eqref{tI3:s} becomes
\begin{equation} \label{tI3:s2}
W_{IJK}\equiv 
 D_I D_{(J} D_{K)} s - \frac{1}{2} \omega_{I(J} D_{K)} \Box s - \frac{1}{4} \omega_{JK} D_{I} \Box s - \omega_{I(J} D_{K)} s + \frac{1}{2} \omega_{JK} D_{I} s = 0.
\end{equation}
It is straightforward to see, by integrating the manifestly non-negative 
$|W_{IJK}|^2$ over the 2-sphere and integrating by parts, that
\be
\int d\Omega\, |W_{IJK}|^2 = -\fft14 \int d\Omega\, s\, \Box\, (\Box+2)
  (\Box + 6)\, s
\ee
and that therefore equation \eqref{tI3:s2} is satisfied if and only if
$s$ is an $\ell=0$, $\ell=1$ or $\ell=2$ spherical harmonic.

In summary, the non-integrable part of $\ndelta \widetilde{\mathcal{I}}_3$ 
vanishes, in general, if and only if $s$ is an $\ell=0$, $\ell=1$ or 
$\ell=2$ spherical harmonic.  In this case, $\ndelta \widetilde{\mathcal{I}}_3$ is integrable, with the charge corresponding to
\begin{equation}
 \widetilde{\mathcal{I}}_3 = - s \ D_I D_J \widetilde{E}^{IJ}.
\end{equation}
This gives a charge that is itself trivially zero for 
$\ell=0$ or $\ell=1$ spherical harmonics,
since they obey $D_I D_J\, s =\ft12 \omega_{IJ}\, \Box\, s$ and $\widetilde E^{IJ}$
is trace-free.  Hence we obtain a non-trivial integrable charge if and only
if $s$ is an $\ell=2$ spherical harmonic. 
This complements the result we obtained in Ref.\ \cite{fakenews}, 
where we found that $\ndelta \mathcal{I}_3$ is integrable and gives a
non-trivial charge if and only if $s$ is an $\ell=2$ spherical harmonic.  As with previously considered subleading charges, we may define a single charge that encapsulates the integrable parts of $\ndelta \mathcal{I}_3$ and $\ndelta \widetilde{\mathcal{I}}_3$ as its real and imaginary parts, respectively~\footnote{Note that there is an unimportant factor of $1/2$ typographical error in equation (3.4) of Ref.\ \cite{fakenews}, which has lead to factor of $1/2$ discrepancies in some other equations, such as equation (4.29) of Ref.\ \cite{fakenews}.}
\begin{equation} \label{NPF:3}
 \mathcal{Q}_3 =  \frac{1}{48\pi G} \int d\Omega\ s\; \bar{\eth}^2 \psi_0^1,
\end{equation}
\begin{equation}
 \mathcal{Q}_3 = \mathcal{Q}_3^{(int)} - i \widetilde{\mathcal{Q}}_3^{(int)},
\end{equation}
where
\begin{gather}
 \mathcal{Q}_3^{(int)} = \frac{1}{16\pi G} \int d\Omega\  s\, D_I D_J \Big( -E^{IJ} + \frac{1}{2}\, \textup{tr}E \ \omega^{IJ} \Big), \notag \\[2mm]
 \widetilde{\mathcal{Q}}_3^{(int)} =  \frac{1}{16\pi G} \int d\Omega\ 
s\ \Big(-  D_I D_J  \widetilde{E}^{IJ}\Big) .
\end{gather}
Now, choosing $s$ to be an $\ell = 2$ spherical harmonic so that the non-integrable parts of $\ndelta \mathcal{I}_3$ and $\ndelta \widetilde{\mathcal{I}}_3$ vanish, from equation \eqref{NPF:3}, we obtain a conserved charge~\footnote{Setting $s = \bar{Y}_{2,m}$ is not strictly correct, because we should really be using a real basis of spherical harmonics, since $s$ is real.  However, it is more convenient to work with a complex basis of spherical harmonics.  Clearly, this choice makes no substantive difference as we could equally think of choosing $s$ to be the real and imaginary parts of $\bar{Y}_{2,m}$. See \cite{fakenews,dual0} 
for a more
extensive discussion of this point.}
\begin{equation}
 \mathcal{Q}^{}_3 |_{s^{(\ell=2)}} =  \frac{1}{48\pi G} \int d\Omega\ \bar{Y}_{2,m}\; \bar{\eth}^2 \psi_0^1.
\end{equation}
Integrating by parts we find that this is indeed equal to the NP charges \cite{NP}
\begin{equation}
 \mathcal{Q}^{}_3 |_{s^{(\ell=2)}} = \mathcal{Q}^{(NP)} = \frac{1}{4\sqrt{6}\pi G} \int d\Omega\ {}_{2}\bar{Y}_{2,m}\; \psi_0^1.
\end{equation}

\section{Discussion} \label{sec:dis}

In this paper, we have resolved two puzzles arising from earlier work \cite{fakenews, dual0}; we have extended the notion of dual gravitational charges to subleading orders in a $1/r$ expansion away from null infinity and found that at the
order $1/r^3$ this, together with the subleading charges proposed in Ref.\ \cite{fakenews}, accounts for all ten of the non-linear Newman-Penrose charges.

The tower of dual gravitational charges is constructed from a new 2-form $\widetilde{H}$ (see equation \eqref{tH}) and can be viewed as being dual to the BMS charges constructed from the Hodge dual of the Barnich-Brandt 2-form $H$ (see equation \eqref{H}).  At the leading order, restricting to supertranslations, the two 2-forms coincide (see equation \eqref{HtH}).  However, they are different at lower orders in a $1/r$ expansion away from null infinity.  

The Barnich-Brandt 2-form is derived by considering the linearised Einstein equation and defining a quantity that vanishes on-shell. The electric charge is the surface integral of a current that is conserved upon use of Maxwell's equation. Analogously the Barnich-Brandt 2-form defines a quantity that is the surface integral of a current that vanishes upon use of the linearised Einstein equation and may be viewed as the analogue of the electric charge.  

On the other hand, we construct the dual charges using a 2-form that is a total derivative (see appendix~\ref{app:charge}). In this sense it is analogous to a magnetic Komar charge and defines a charge without use of the Einstein equation. However, nevertheless we obtain charges that are non-trivial and account for the recently proposed dual gravitational charges \cite{dual0}, extending them to charges associated with the full BMS group, and the imaginary parts of the non-linear Newman-Penrose charges. 

\section*{Acknowledgements}

M.G.\ is partially supported by grant no.\ 615203 from the European Research Council under the FP7.  C.N.P.\ is partially supported by DOE grant DE-FG02-13ER42020.

\appendix

\section{Boundary terms} \label{app:charge}
In this section, we prove that the variation of the dual charge \eqref{Dcharge} vanishes, using the fact that
\begin{equation} \label{metvar}
 \delta g_{ab} = 2 \nabla_{(a} \xi_{b)}.
\end{equation}
From the definition of the dual 2-form $\widetilde{H}$, given in equation \eqref{tH}, and using equation \eqref{metvar}
\begin{align}
2 \widetilde{H}_{IJ} &= \xi^c \nabla_J \nabla_I \xi_c + \xi^c \nabla_J \nabla_c \xi_I + \nabla_J \xi^c \nabla_c \xi_I \notag \\
                 &= R_{JIcd} \xi^c \xi^d + \nabla_J (\xi^c \nabla_c \xi_I) \notag \\
                 &= \partial_J (\xi^c \nabla_c \xi_I),
\end{align}
where we assume an antisymmetrisation in $[IJ]$ on the right hand side for each equality.  In the second equality we use the fact that
\begin{equation}
 [\nabla_a,\nabla_b] V_c = R_{abc}{}^d V_d.
\end{equation}
Thus, we conclude that $\ndelta \widetilde{\mathcal{Q}}$ as defined in \eqref{Dcharge} vanishes.

\section{Further properties of $\widetilde{H}$} \label{app:Htilde}
In this section, we verify that 
\begin{equation}
 \ndelta \widetilde{\mathcal{Q}}[\xi, g, \delta g = \mathcal{L}_\zeta g] = -\ndelta \widetilde{\mathcal{Q}}[\zeta, g, \delta g = \mathcal{L}_\xi g] 
\end{equation}
as one would expect.

Starting from equation \eqref{tH},
\be
\widetilde H[\xi, g, \delta g] = \ft12 \Big\{\xi^c\, \nabla_b\, \delta g_{ac} - 
   \ft12\delta g_{bc}\,(\nabla_a\xi^c-\nabla^c\xi_a)\Big\}\, dx^a\wedge dx^b\,,
\ee
we can write the first term as $\nabla_b(\xi^c\, \delta g_{ac}) -
  \delta g_{ac}\, \nabla_b\xi^c$, and hence we get
\be
\widetilde H[\xi, g, \delta g] = \ft12 \Big\{\nabla_b(\xi^c\, \delta g_{ac}) +
         \ft12 \delta g_{bc}\, (\nabla_a\xi^c+\nabla^c\xi_a)\Big\}\,
  dx^a\wedge dx^b\,.
\ee
Thus we have
\be
\widetilde H[\xi, g, \delta g] = -\ft12 d(\xi^c\, \delta g_{ac}\, dx^a) +
  \ft14 \delta g_{bc}\, g^{cd}\, (\nabla_a\xi_d+\nabla_d\xi_a)\, dx^a\wedge dx^b
\,.\label{H2}
\ee
If we take $\delta g_{bc}$ to be a variation $\delta_{\zeta}\, g_{bc}$ coming from
a BMS generator $\zeta$ with $\delta_{\zeta}\, g_{bc}=\nabla_b\zeta_{c} +
  \nabla_c\zeta_{b}$, and view $(\nabla_a\xi_{d}+\nabla_d\xi_{a})$ as defining a 
metric variation $\delta_\xi\, g_{ad}$, then we have
\be
\widetilde H[\xi, g, \delta g] = d\omega + \ft14 (\delta_\xi\, g_{ac})\, (\delta_\zeta\, g_{bd})\,
   g^{cd}\, dx^a\wedge dx^b\,,
\ee
where $\omega =-\ft12 d(\xi_2^c\, \delta_1\, g_{ac}\, dx^a)$.  

Equation \eqref{Dcharge} defines $\ndelta \widetilde{\mathcal{Q}}[\xi, g, \delta g]$ as the integral of $\widetilde H[\xi, g, \delta g]$.  Thus, we conclude that
\begin{equation}
 \ndelta \widetilde{\mathcal{Q}}[\xi, g, \delta g = \mathcal{L}_\zeta g] = -\ndelta \widetilde{\mathcal{Q}}[\zeta, g, \delta g = \mathcal{L}_\xi g].
\end{equation}

\bibliographystyle{utphys}
\bibliography{NP}

\end{document}